\documentstyle[12pt,fleqn]{article}
\input epsf

\def\singlespace {\smallskipamount=3.75pt plus1pt minus1pt
                  \medskipamount=7.5pt plus2pt minus2pt
                  \bigskipamount=15pt plus4pt minus4pt
                  \normalbaselineskip=15pt plus0pt minus0pt
                  \normallineskip=1pt
                  \normallineskiplimit=0pt
                  \jot=3.75pt
                  {\def\smallskip {\vskip\smallskipamount}}
                  {\def\medskip   {\vskip\medskipamount}}
                  {\def\bigskip   {\vskip\bigskipamount}}
                  {\setbox\strutbox=\hbox{\vrule 
                    height10.5pt depth4.5pt width 0pt}}
                  \parskip 7.5pt
                  \normalbaselines}
\def\middlespace {\smallskipamount=5.625pt plus1.5pt minus1.5pt
                  \medskipamount=11.25pt plus3pt minus3pt
                  \bigskipamount=22.5pt plus6pt minus6pt
                  \normalbaselineskip=22.5pt plus0pt minus0pt
                  \normallineskip=1pt
                  \normallineskiplimit=0pt
                  \jot=5.625pt
                  {\def\smallskip {\vskip\smallskipamount}}
                  {\def\medskip   {\vskip\medskipamount}}
                  {\def\bigskip   {\vskip\bigskipamount}}
                  {\setbox\strutbox=\hbox{\vrule 
                    height15.75pt depth6.75pt width 0pt}}
                  \parskip 11.25pt
                  \normalbaselines}
\def\doublespace {\smallskipamount=7.5pt plus2pt minus2pt
                  \medskipamount=15pt plus4pt minus4pt
                  \bigskipamount=30pt plus8pt minus8pt
                  \normalbaselineskip=30pt plus0pt minus0pt
                  \normallineskip=2pt
                  \normallineskiplimit=0pt
                  \jot=7.5pt
                  {\def\smallskip {\vskip\smallskipamount}}
                  {\def\medskip   {\vskip\medskipamount}}
                  {\def\bigskip   {\vskip\bigskipamount}}
                  {\setbox\strutbox=\hbox{\vrule 
                    height21.0pt depth9.0pt width 0pt}}
                  \parskip 15.0pt
                  \normalbaselines}

\def\be{\begin{equation}}
\def\ee{\end{equation}}

\def\sect #1{\setcounter{equation}{0}}

\begin{document}

\singlespace

\begin{center}
{\Large {The final fate of spherical inhomogeneous dust collapse II: 
{\Large {Initial data and causal structure of singularity}}}}
\end{center}
\vspace{1.0in}
\vspace{12pt}
\begin{center}
{\large{Sanjay Jhingan, P. S. Joshi and T. P. Singh\\
Theoretical Astrophysics Group\\
Tata Institute of Fundamental Research\\
Homi Bhabha Road, Bombay 400005, India.\\
}}
\end{center}
\vskip 1 in
\centerline{\bf ABSTRACT}

\medskip

\noindent Further to results in [9], pointing out the role of initial
density and velocity distributions towards determining the final outcome
of spherical dust collapse, the causal structure of singularity is examined 
here in terms of evolution of the apparent horizon. We also bring out several 
related features which throw some useful light towards understanding the 
nature of this singularity, including the behaviour of geodesic families 
coming out and some aspects related to the stability of singularity.

\eject

\noindent {\bf I. Introduction}

\smallskip

The spherical gravitational collapse of inhomogeneous dust has been
analyzed in several recent papers [1-9]. We now have a reasonably good
understanding of the role of initial data in deciding whether the
central singularity is naked or covered [8,9]. Continuing the analysis 
of [9], in the present note, we discuss the relation between the initial 
data and the dynamics of the apparent horizon - this study improves our 
understanding of the connection between initial data and the nature of the 
final singularity. 

In Section II we recall some essential results and discuss 
some aspects of initial data which provide certain useful insights into 
the structure of the singularity. In particular, we consider the stability 
of naked singularity under perturbation of initial data. The dynamics of 
evolving apparent horizon is discussed in Section III for different initial 
data and Section IV examines in some detail the behaviour of outgoing 
families of geodesics, when the singularity is visible, for different 
classes of initial data.

Our overall results in this paper are as follows. We argue that the strong
naked singularities in dust collapse are non-generic,
in a certain sense discussed below. We show that the
Oppenheimer-Snyder solution is stable against perturbations leading to
a strong naked singularity, but unstable against perturbations leading to
a weak naked singularity. We study the dynamics of the apparent horizon 
and show that the evolution of the apparent 
horizon for initial data leading to a naked singularity is very different 
from the case when the initial conditions lead to a covered singularity.
In particular, we provide an example to show that absence of the apparent
horizon until singularity formation does not imply the singularity is naked.
This has relevance to searches for naked singularities in numerical 
investigations of collapse. We show that the geodesic structure for the
weak naked singularity is different from that for the strong naked
singularity.

\bigskip

\noindent {\bf II. Initial data and nature of the singularity}

\smallskip

Spherical dust collapse is described by the Tolman-Bondi model [10,11] 
in comoving coordinates $(t,r,\theta,\phi)$ which has metric of the form
\be
ds^2 = -dt^2 + {R'^{2}\over 1+f(r)} dr^2 + R^2 d\Omega^2,
\ee
and energy-momentum tensor of the form of a perfect fluid with
equation of state p=0, i.e., ${T^{ij} = \epsilon \,
\delta^{i}_{t}\delta^{j}_{t}}$.
Here $\epsilon$ and $R$ are functions of $r$ and $t$,
$d\Omega^2$ is the metric on the two-sphere $S^2$, and $f(r)$ is an arbitrary
function of $r$ which we call the energy function. The Einstein
equations are
\be
{\dot R}^2 = \frac{F(r)}{R}  + f(r), 
\ee
\be
\epsilon(r,t)= \frac{F'(r)}{R^2 R'},
\ee
where the dot and prime denote $\partial/ \partial t$ and 
$\partial/ \partial r$ respectively. $F(r)$ is another free
function of integration, and it is equal to twice the mass inside the sphere
of radius $r$.

As we are concerned with the situation of gravitational collapse, we 
take $\dot R<0$. The evolution leads to the formation of a shell-focussing 
curvature singularity described by the curve $R(t,r)=0$. The singularity 
at $r=0$ is called the central singularity. The free functions $F(r)$ and 
$f(r)$ are determined by the initial density profile $\rho(r)=\epsilon(r,0)$ 
and the initial velocity profile $v(r)=\dot R(r,0)$
as follows. Let $t=0$ be the initial epoch, and by virtue of the
freedom of scaling let us choose $R(0,r) = r$. Then we get from (3),
\be
F(r) = \int \rho(r) r^{2} dr,   
\ee
and from (2) and (4) that
\be
f(r) = v^{2}(r) -   {1\over r} \int \rho(r) r^{2} dr.
\ee 
We assume that the initial density and velocity profiles can be expanded
in a power series, near the center $r=0$,
\be
\rho (r) = \rho_{0} + \rho_{1}r + {1\over 2}\rho_{2}r^{2} + 
{1\over 6} \rho_{3}r^{3} + ...
\ee
\be
v(r) =  v_{1}r + {1\over 2} v_{2}r^{2} + {1\over 6}v_{3}r^{3} + ...
\ee
We assume the density to decrease outwards from the center, hence the
first non-vanishing derivative of density has negative sign. The terms 
$\rho_{n}$ and $v_{n}$ denote the $n\it{th}$ derivative at the center, of the
respective quantity. The inclusion of the $\rho_{1}$ term in (6) means
there is a central cusp in the initial density, and such a term should be 
dropped to avoid this feature. It is included here for completeness, and 
results have been worked out both with and without this term. In (7), the 
first term in the expansion is of order $r$ - the center is taken to be at 
rest, because of spherical symmetry. From (4) and (6) it follows that the 
mass $F(r)$ near the center is given by
\be
 F(r) = F_{0}r^{3} + F_{1}r^{4} + F_{2}r^{5} + F_{3}r^{6} +  ... 
\ee
where $F_{n} = \rho_{n}/(n+3)n!$.
Using (5), (7) and (8) it is seen that the energy function $f(r)$ is
given near the center by 
\be
f(r) = f_{2}r^{2}  +  f_{3}r^{3} + f_{4}r^{4}  + f_{5}r^{5} + ...
\ee
where the coefficients $f_{n}$ are determined by the coefficients in
(7) and (8). Note that $f(r)$ cannot have a term lower than the
quadratic one near the center - this is a consequence of the center being
at rest, and of the fact that $F(0)$, the mass at the center, is zero.
Thus $f(r)$ must necessarily vanish at $r=0$. Further, we assume 
$f_{2}\neq 0$, as its vanishing is only possible by a fine tuning of the
density and velocity profiles. 

It can be shown that the shell-focussing singularity at $r>0$ is covered 
by the event horizon, but the central singularity at $r=0$ is locally naked 
for some initial data, and covered for other initial data. (In this paper, 
when we say the singularity is naked, we mean it is locally naked. We will 
not be concerned with the issue of global nakedness which has been discussed
elsewhere [7,9]). Also, the results on the curvature strength of the naked 
singularity, as to whether it is strong or weak (this is a technical
criterion quantifying the rate of curvature growth in the limit of approach
to the singularity along the outgoing future directed nonspacelike 
trajectories; see e.g. ref.[7] for definitions), have been worked out. 
For the marginally bound case, $f=0$, these results are as follows [3,6,9]:

\noindent (i) If $\rho_{1}<0$ the singularity is naked and weak.

\noindent (ii) If $\rho_{1}=0,\rho_{2}<0$ the singularity is naked and weak.

\noindent (iii) If $\rho_{1}=\rho_{2}=0,\rho_{3}<0$ the singularity is
naked if $\xi=F_{3}/F_{0}^{5/2} =\sqrt{3}\rho_{3}/4\rho_{0}^{5/2}$ is less 
than the critical value $\xi_{c}=-25.9904$, and covered if $\xi>\xi_{c}$. 
Further, the naked singularity is a strong curvature singularity.

\noindent (iv) If $\rho_{1}=\rho_{2}=\rho_{3}=0$ the singularity is covered.  

For the non-marginally bound case $f\neq 0$ the results are as follows. We
define the quantity $Q_{q}$ as
\be
 Q_{q} ={3\over 2} 
 \left(1-{f_{2}\over 2F_{0}}\right)
 \left( G(-f_{2}/F_{0})\left({F_{q}\over F_{0}}-{3f_{q+2}\over 2f_{2}}\right)
\left(1+{f_{2}\over 2F_{0}}\right) + {f_{q+2}\over f_{2}}- {F_{q}\over F_{0}}
\right).
\ee
Here $q=1$ if at least one of $F_{1}$ and $f_{3}$ are non-zero, $q=2$ if
$F_{1}=f_{3}=0$ and at least one of $F_{2}$ and $f_{4}$ are non-zero, and
$q=3$ if $F_{1}$, $F_{2}$, $f_{3}$ and $f_{4}$ are zero, and at least one
of $F_{3}$ and $f_{5}$ are non-zero. If $q=1$, the singularity is naked 
for $Q_{1}>0$ and if naked it is weak. If $q=2$ the singularity is naked 
for $Q_{2}>0$ and if naked it is weak. If $q=3$ the singularity is naked if 
$\xi=-2Q_{3}/F_{0}^{3/2}$ is less than the above mentioned critical value 
$\xi_{c}$, and covered if $\xi$ exceeds $\xi_{c}$. Further the naked 
singularity is strong. If all of $F_{1}$, $F_{2}$, $F_{3}$, $f_{3}$, $f_{4}$ 
and $f_{5}$ are zero, the singularity is covered. 

We now make several observations here regarding the nature of 
this singularity, which arises as the final state of inhomogeneous 
gravitational collapse of pressureless dust, and its relationship with
the initial data. In our view, this brings out and throws some useful light 
on certain important aspects related to this singularity which are 
not widely known and which should help us understand better the structure
of singularity.  

To begin with, we should comment on the series expansions in Eqns. (6) to (9).
In their work, Christodoulou [2] and Newman [3] assumed that the density
and metric functions are smooth ($C^{\infty}$). It then follows that a
power series expansion near the center for the density $\rho(r)$ and 
for the energy function $f(r)$ can have only even powers of $r$. This
is because if odd powers of $r$ are present, then in a Cartesian 
coordinate system set up at the center, some of the Cartesian derivatives 
of odd-powered terms are not defined - this makes the function non-smooth.
For instance, if in (6), the linear term is absent, and the cubic term
present, the density function is $C^{2}$, but not $C^{\infty}$.
Further, since the spherical coordinate system is singular at the origin,
strictly one must set up Cartesian coordinates to describe quantities near
the center.

As is evident from Eqns. (6) to (9), we are not restricting ourselves to
smooth functions. It could be asked, does a physical density function have
to be smooth [12], where by physical we mean `occuring in a real system'.
In our view, the answer is in the negative. As a justification and
for illustration, we consider the case of a spherically symmetric star in
Newtonian gravity, described by a polytropic equation of state. If the
star is in hydrostatic equilibrium, the density distribution is described
by the Lane-Emden equation [13], and the substitution of a power-series
expansion for the density into the equation indeed implies that all 
odd-powered terms drop out and the density function is necessarily smooth. 
However, if one is considering gravitational collapse of the star, the 
Lane-Emden equation is replaced by the dynamical equation for the evolution 
of the radius $R$ of a fluid element,
$$  \ddot{R} = -{4\pi\over R^{2}}\int \rho R^{2} dR - 
    {1\over \rho} {dp\over dR},  $$
where $p=K\rho^{1 + 1/n}$ describes the polytrope. Setting $\ddot{R}=0$ 
reduces this equation to the Lane-Emden equation. While
considering collapse, initial conditions must be set using the above
dynamical equation and {\it not} using the Lane-Emden equation. Thus when
a power-series expansion for the initial density, having odd as well as
even powers, is substituted in this dynamical equation, the odd powers do
not drop out. Instead, this equation determines the initial acceleration,
given an initial density profile. A situation analogous to this Newtonian
case holds in general relativity as well. It should also be emphasized that
the restriction to even powers arising from the Lane-Emden equation is not
applicable to dust collapse as such, there being no pressure, and hence no
equilibrium.

We conclude that a priori, physical density functions need not necessarily
be smooth functions. (See also, for instance, the discussion in [14] of
observations suggesting possible cusps in globular cluster cores).
In certain equilibrium cases, the field equations
imply that they have to be smooth, but this is not true in general. Also, 
there need not be a restriction on the initial velocity distribution to be 
smooth. In our view, physical quantities should at the most be required to
be $C^{2}$, so as to ensure solvability of the field equations. For 
instance, the {\it self-similar} density profile, which is often of 
physical interest (see e.g. [15], [16]), is a $C^{2}$ function in the case of 
dust [16]. 

All the same, it is important to note that the issue of whether 
physical quantities should be smooth or not does not have a bearing on our 
overall conclusion regarding 
censorship in dust collapse. As we explain below, the occurence of a strong
naked singularity does not necessarily require the density to be non-smooth;
it is sufficient that the density is smooth, and the energy function $f(r)$
is $C^{4}$, (the velocity $v(r)$ is then $C^{2})$. However, the strong naked
singularities are non-generic in either case, in the sense described below.
The generic naked singularity is weak and hence it is {\it genericity}, 
as opposed to {\it smoothness}, which limits the importance of strong naked
singularities. In other words, even if
one allows non-smoothness in the initial data, any resulting strong naked
singularities are non-generic. Perhaps it will also be useful to mention that
when one considers equations of state other than dust, strong naked 
singularities do arise from smooth initial data [16].

As pointed out above, in the marginally bound case, the strong 
singularity arises from an initial density profile involving a cubic term, 
which is not smooth, but $C^{2}$. (We wish to recall that when 
we call a function of $r$ non-smooth we mean that some of the Cartesian
derivatives are ill-defined; all the derivatives with respect to $r$ are 
well-defined though). However, an important question in this
connection is the following: Is it true that a strong curvature naked 
singularity must necessarily involve and arise from a non-smooth density 
profile? The answer to this question is no. 
In fact, as can be deduced from the above, in the non-marginal case,
one can get a strong naked singularity from density functions that are
smooth, and a metric that is $C^{4}$. Consider, for instance, the
case $\rho(r)=\rho_{0}=$const., or a smooth $\rho(r)$ with $\rho_{2}=0$, and 
$f_{3}=f_{4}=0$, $f_{5}<0$. This initial data again leads to a strong naked 
singularity. Hence, non-smoothness of the density is not a pre-requisite
for the strong naked singularity to occur.

Next, let us consider the stability of the Oppenheimer-Snyder (O-S) 
dust collapse solution [17] (as a special case of models considered here), 
under perturbation of initial data, in 
the marginally bound collapse. As we know, the O-S model describes homogeneous
dust collapse leading to the formation of a black hole. If the O-S initial 
data is perturbed by switching on an infinitesimal negative $\rho_{1}$ or 
$\rho_{2}$ terms, we get a naked singularity instead. The black hole is 
not stable 
to small perturbations. On the other hand, if the O-S black hole scenario
is perturbed 
by switching on a small $\rho_{3}$, it continues to be a black hole. Only 
large enough perturbations from homogeneity at the level of the 
third derivative convert the black hole to a naked singularity.
If we make the plausible proposition that the O-S black hole must be stable
to small perturbations in density, we can conclude that the weak naked 
singularity is not a genuine naked singularity. On the other hand, from this
point of view,
the strong naked singularity is a genuine naked singularity. This inference 
is strengthened by the likely possibility that spacetime may be extendible 
through a weak enough naked singularity, but not through a strong naked 
singularity.

We now comment on the possible measure of initial data which leads 
to a strong naked singularity in the case of inhomogeneous dust collapse. 
This argument is due to Reza Tavakol.
Consider first the marginally bound case. Only the three derivatives 
$\rho_{1},\rho_{2}$ and $\rho_{3}$ play a role in deciding whether or not 
the singularity is naked. Hence consider a three-dimensional space of initial 
data, labelled by these three derivatives. A generic point in this space 
will have all the three derivatives non-zero. For physical reasons we 
restrict to the half-space $\rho_{1}<0$. Since the first non-vanishing 
derivative is the relevant one, it follows that the generic singularity is 
weak and naked. If we set $\rho_{1}=0$ the generic naked singularity continues
to be weak and naked. The strong naked singularity arises from part of the 
half-line $\rho_{1}=\rho_{2}=0,\rho_{3}<0$, and hence from initial data that 
is of measure zero in this three-space; (see also [18]).
A similar consideration holds for the
non-marginal case - the relevant space is now six dimensional, consisting
of the first three derivatives of $\rho(r)$ and of $f(r)$. It can be
concluded that the strong naked singularity arises from an initial data
set of measure zero. Perhaps it should be emphasized that this result
(strong naked singularity arises from data of zero measure)
is specific to dust collapse, and may not hold for more general forms
of matter. In fact, in the gravitational collapse of imploding radiation
described by the Vaidya space-time, an initial data set of non-zero
measure leads to a {\it strong} naked singularity [19]. In this model no 
initial data lead to a {\it weak} naked singularity. 

The preceding two arguments together suggest that the generic  
naked singularity in dust collapse is weak, which we are proposing 
is not a genuine singularity. Should this be taken to mean that dust collapse 
is consistent with the cosmic censorship hypothesis? According to us the 
answer is no. Even if the singularity is weak (the quantity $R_{ij}V^iV^j$ 
grows as $1/k$, rather than as $1/k^2$ for the case of a strong 
singularity, where
$k=0$ at the center) and spacetime is possibly extendible, one will still 
see regions of arbitrarily high curvatures without any bound via the outgoing 
nonspacelike geodesics which start from near the center. As we see it, while
the above arguments provide a good pointer to a possible direction of a 
cosmic censorship statement, their rigorous mathematical formulation and
proof might turn out to be a difficult task to achieve as has been the case
for the attempts so far. This is mainly because of the formidable difficulties
in formulating a suitable stability analysis in general relativity and the
related issues such as the complexity caused by the non-linearity of Einstein 
equations etc. What is needed, in our view, is some form of physical 
formulation and justification of the cosmic censorship principle [20]. 

\bigskip

\noindent{\bf III. Initial data and dynamics of the apparent horizon}.

\smallskip

The behaviour and evolution of the apparent horizon in the course
of gravitational collapse gives insight into the causal structure near
the singularity. In this Section we work out some aspects of the
evolution of the apparent horizon for the marginally bound case, $f=0$ and
examine the relation between initial data and the formation or otherwise 
of a naked singularity. This helps us understand why some initial data 
lead to a naked singularity, while the other would produce a black hole
as the end product of collapse. 
From Eqn. (2), using the scaling $R(0,r)=r$ at the initial epoch $t=0$,
we find the solution to be
\be
      R^{3/2}   = r^{3/2}  -{3\over 2}\sqrt{F} t.
\ee
The singularity curve $R(t_{s}(r),r)=0$ is given by 
$t_{s}(r)=2r^{3/2}/3\sqrt{F}$. The apparent horizon, which is the boundary
of the trapped region, is given by the curve $R(t_{ah}(r),r)=F(r)$, and
using this relation, we find from (11) that
\be
 t_{ah}(r)    =   {2r^{3/2}\over 3\sqrt{F}} - {2\over 3}F(r).
\ee
Evidently, dust shells with $r>0$ become trapped before they become
singular, whereas at $r=0$ the singularity and apparent horizon form
simultaneously. We are interested in the properties of the function
$t_{ah}(r)$. Although the function $t_{s}(r)$ is monotonically increasing,
$t_{ah}(r)$ is not necessarily monotonic. 

Let us first work out its behaviour for
the Oppenheimer-Snyder model. Now, we have $F(r)=F_{0}r^{3}$ and hence
\be
      t_{ah}(r)  =  {2\over 3\sqrt{F_{0}}}  - {2F_{0}\over 3}r^{3}.
\ee
Clearly, the boundary $r_{b}$ of the star gets trapped first, and the 
apparent horizon moves inwards, the center being the last point to get 
trapped. The event horizon begins to form a finite time before the boundary
gets trapped. (See Fig. 1(i)).
Note that for the homogeneous case, $t_{ah}(r)<t_{ah}(0)$ near the origin,
signifying that the neighborhood of the center gets trapped before the
center. Since $t_{ah}(0)=t_{s}(0)$ the neighboring regions get trapped
before the center becomes singular, and this helps understand why the central
singularity is not naked.

Consider now the case of inhomogeneous collapse. The global evolution of
the apparent horizon will depend on the nature of $F(r)$ throughout the star, 
but near
$r=0$ it will be determined by $F(r)$ near the center. If at $r=0$
the first non-vanishing derivative of the density is the $n{\it th}$ one,
then using (8) and (12) we get to leading order
\be 
   t_{ah}(r)   =   {2\over 3\sqrt{F_{0}}}
                   -{F_{n}r^{n}\over 3F_{0}^{3/2}}
                   -{2\over 3}F_{0}r^{3}.
\ee

This equation helps us understand the evolution of the apparent horizon
for an inhomogeneous density profile and the following conclusions can be
drawn. If the first non-vanishing derivative of the density at the center 
is either $\rho_{1}$ or $\rho_{2}$, the so called weak naked singularity 
occurs and in that case $t_{ah}(r)>t_{ah}(0)=t_{s}(0)$. That is, the center 
becomes singular before its neighborhood gets trapped and the first point
of singularity coincides with the first point in time of the apparent horizon.
The qualitative behaviour of the apparent horizon formation is as shown 
in Fig. 1(ii). The dynamical evolution of apparent horizon now is totally 
different from the homogeneous case, and helps 
understand why in this case the singularity is naked.

Next, if the first non-vanishing derivative is $\rho_{3}$, we get
$t_{ah}(r) > t_{ah}(0)$ if the parameter $\xi$ defined earlier satisfies
$\xi<-2$ and $t_{ah}(r)<t_{ah}(0)$ if $\xi>-2$. Recall that the
singularity is naked for $\xi<-25.9904$ and covered if $\xi$ is greater
than this number. This means there is the range $-25.9904<\xi<-2$ in which 
the singularity is not naked, even though the center gets trapped before
its neighborhood. Put differently, the apparent horizon is absent until
the formation of the singularity, but the singularity is not naked. This
is a counterexample to the criterion used by Shapiro and Teukolsky in their
numerical studies [21], where they suggest that if the apparent horizon is
absent until singularity formation, the singularity is naked. As is clear
from this consideration, the condition
that the center get trapped before its neighboring region is necessary but
not sufficient for nakedness.

\begin{center}
\leavevmode\epsfysize=3.5 in\epsfbox{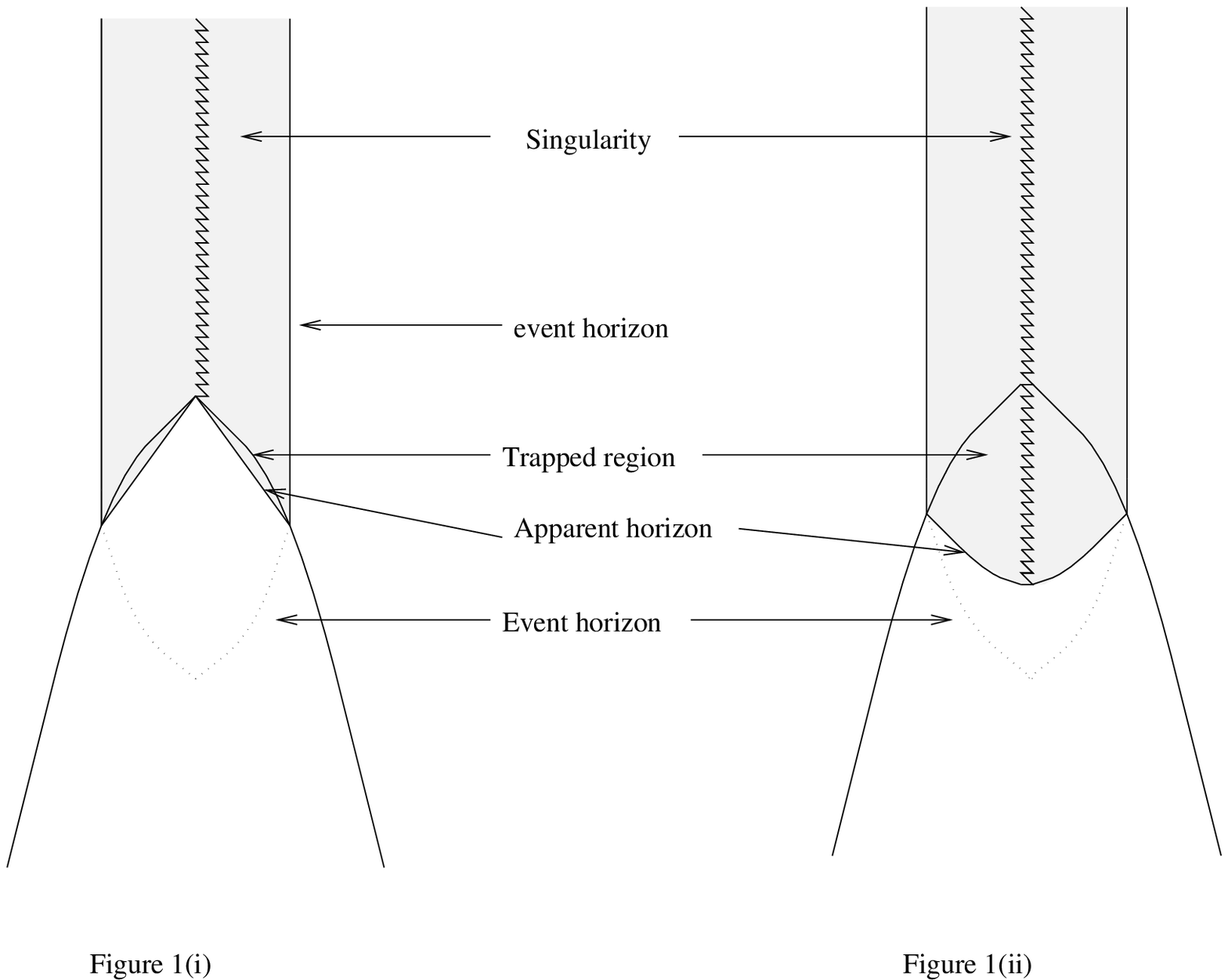}
\end{center}
\noindent {\small Fig. 1(i): The standard picture of spherical homogeneous dust
collapse, the Oppenheimer-Snyder model. The shaded portion is the trapped
region. Note that first the boundary of the star gets trapped, then the
trapped region expands into the star, towards the center.
The center is the last point to get trapped. The curvature singularity
forms simultaneously all over the star. Fig. 1(ii): The corresponding 
space-time diagram for
inhomogeneous dust-collapse leading to a locally naked singularity. The
apparent horizon curve is as given by Eqn. (16), with $\xi < -2$. The
shaded portion is the trapped region. Note that unlike in the 
Oppenheimer-Snyder model, the center gets trapped first, and the trapped
region moves out towards the boundary of the star. Different shells
become singular at different times, the center becoming singular first.}

\bigskip

If the first three derivatives are zero, then $n\geq 4$ and it follows
from (14) that $t_{ah}(r)<t_{ah}(0)$ - the center gets trapped later
than its neighborhood, and hence the central singularity is not naked;
the qualitative picture being again same as Fig. 1a. We
see that there is a qualitative change in the behaviour of the apparent
horizon near the center as the density profile (6) is made more and more
homogeneous by setting successive derivatives to zero. This provides a
physical picture for the relation between the initial data and the nature
of the singularity, as to whether it is naked or covered.

As is clear from (12), the global behaviour of the function $t_{ah}(r)$
can be worked out only if we know $F(r)$ throughout the matter cloud. 
To understand the overall evolution of the apparent horizon within the
cloud, we consider below an illustrative example of a typical initial density 
profile where the density is inhomogeneous, decreasing away from the center,
and we work out the evolution of the apparent horizon through the star. 
Consider the initial profile
\be     
    \rho(r) = \rho_{0}\left( 1 - {r^{3}\over r_{b}^{3}}\right)
\ee
where $\rho_{0}$ is the central density and $r_{b}$ is the boundary of the
cloud. Recall that we are using the scaling $R=r$ at the initial epoch, so 
this is the physical initial density function. We can compute the resulting
mass-function $F(r)$ using (4) and the apparent horizon curve using (12).
We get
\be
  t_{ah}(r)= {2\over \sqrt{3\rho_{0}}}
   \left(1-{r^{3}\over 2r_{b}^{3}}\right)^{-1/2}
   -{2\rho_{0}\over 9}\left(r^{3}- {r^{6}\over 2r_{b}^{3}}\right).
\ee
It can be verified that for $\xi<-2$ this is a monotonically increasing
function. That is, the center is the first point to get trapped, and
shells with larger and larger initial radii get trapped at later and
later times. Again, this behaviour should be contrasted with that of the
trapped surface in the Oppenheimer-Snyder model (see Fig. 1(ii)). 
If $\xi>-2$ the above $t_{ah}(r)$ starts to decrease near the center,
but has a turning point at some location in the interior of the star (other
than the center). This means some point other than the center or the
boundary is the first one to get trapped, and as time progresses, the
trapped region moves outwards as well as inwards. Figs. 2(i) and 2(ii)
show $t_{ah}(r)$
as given by (16), for various representative values of $\xi$. 
In Fig. 2(i) we have set $\rho_0=1$ and varied $r_b$, while in Fig. 2(ii)
we have set $r_b=1$ and varied $\rho_0$.

\begin{center}
\leavevmode\epsfysize=4 in\epsfbox{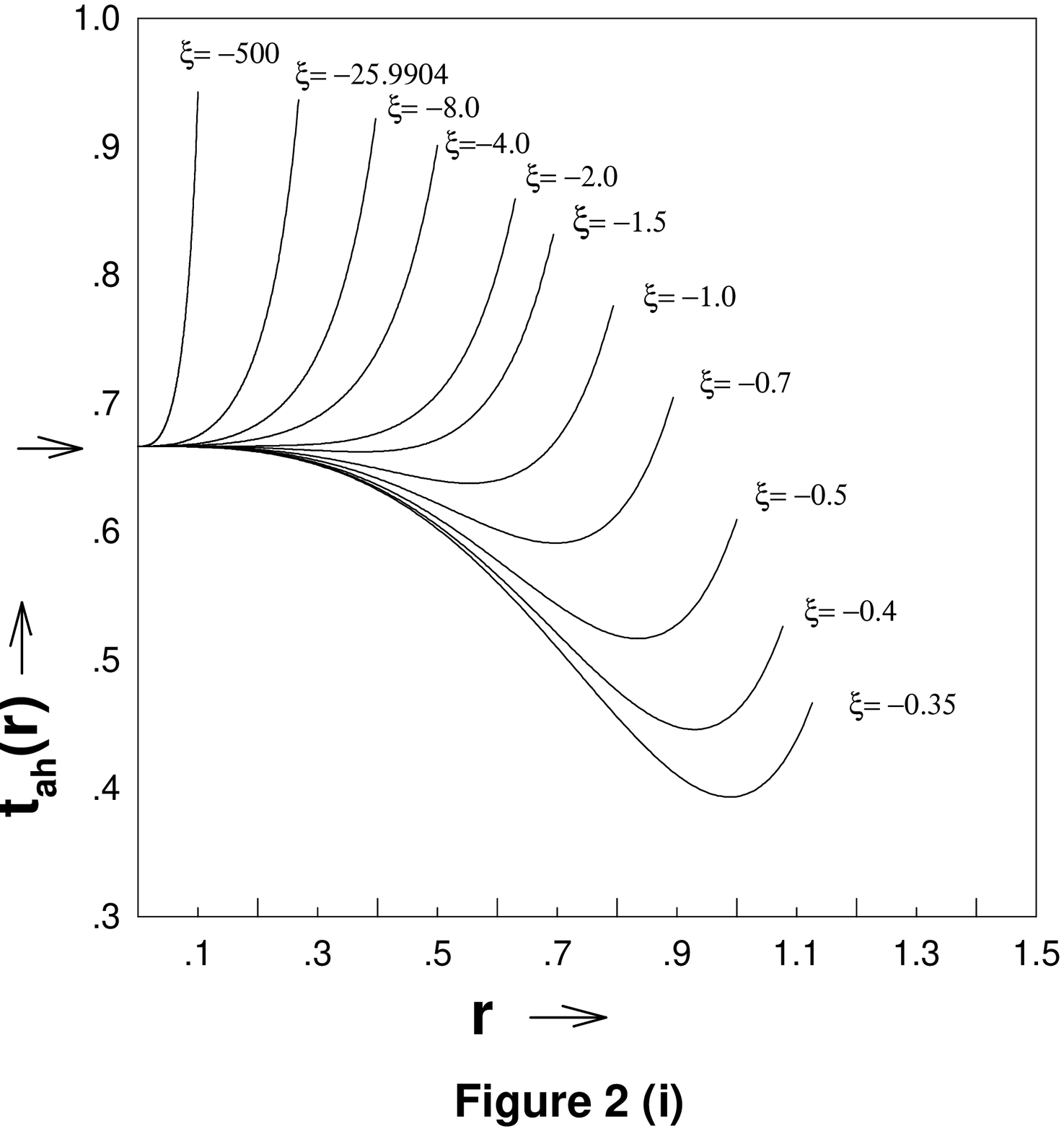}
\end{center}
\vfil\eject

\begin{center}
\leavevmode\epsfysize=4 in\epsfbox{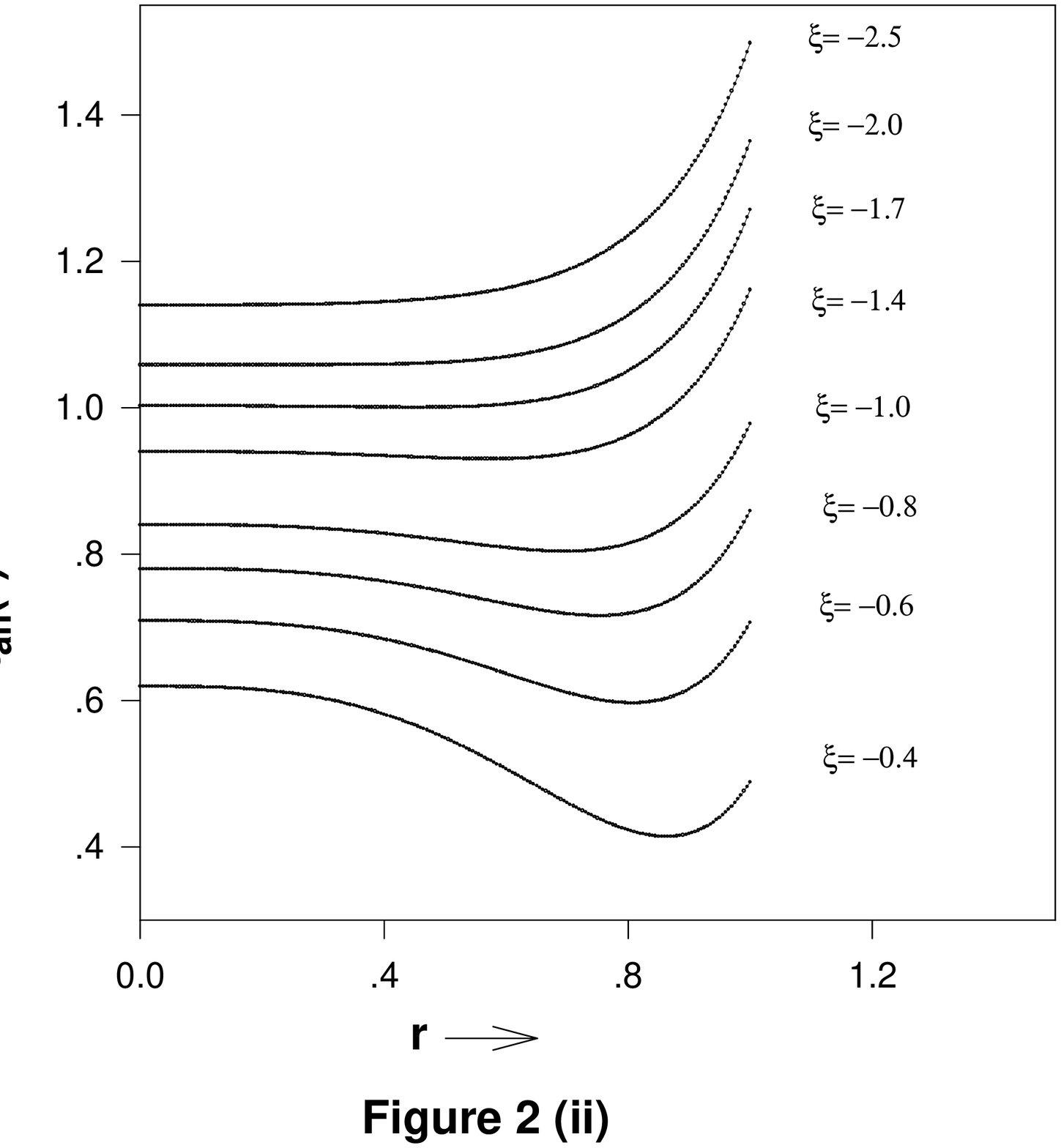}
\end{center}
\noindent {\small Figure 2: A plot of the apparent horizon curves described by
Eqn. (16), for various values of $\xi$. For $\xi<-2$ the center is the
first point to get trapped, whereas for $\xi>-2$ some surface in the
interior of the star is the first one to get trapped, and the trapped
region moves both inwards and outwards.
Fig. 2(i) is obtained by setting $\rho_0=1$ and varying $r_b$, 
whereas in Fig. 2(ii) we have $r_b=1$ and $\rho_0$ has been varied.}

\bigskip

The causal nature of the apparent horizon near the center can be worked
out by writing the induced metric on the apparent horizon. It is also
convenient to use the mass-function $F(r)$ as a coordinate, instead
of $r$. Using (1), (11) and (12) the induced metric is shown to be
\be
     ds^{2} =  \left( 2{\partial t_{s}(r)\over \partial F(r)}
               - {1\over 3}\right) dF^{2}(r). 
\ee
It can then be shown that if the first non-vanishing derivative of the density
at the center is either $\rho_{1}$ or $\rho_{2}$, the apparent horizon
is spacelike, and as we know, the singularity is naked. If the first 
non-vanishing derivative is $\rho_{3}$, the apparent horizon is spacelike 
for $\xi < -0.5$ and past-timelike for $\xi > -0.5$. That is, if the
singularity is naked the horizon is spacelike, and if it is covered, the
horizon is spacelike or past-timelike. If the first non-vanishing derivative
is the fourth or higher, the apparent horizon is past-timelike, and the 
singularity is covered.

The event horizon must be outside the apparent horizon covering the trapped
region. Hence in the homogeneous case it must start forming  
before the first point of the singularity forms at the center. When the
initial density is inhomogeneous and the resulting singularity covered, the 
event horizon will again start forming before the central singularity occurs.
In the case of inhomogeneous collapse leading to a naked singularity, the
event horizon may begin to form prior to or simultaneously with the central
singularity, depending on whether the singularity is only locally naked,
or globally naked as well.

\bigskip

\noindent {\bf IV. Naked singularity and the nature of geodesics}

\smallskip

In this Section we discuss the nature of null geodesics families 
emerging from the naked singularity, for marginally bound collapse. 
In particular, we wish to highlight the structural difference between the 
cases of {\it weak} and {\it strong} naked singularities. 
In order to avoid some repetition of earlier results, we refer to [7] for 
the derivation of the geodesic equation in this case. 
We describe the evolution of geodesics in terms of the pair of 
variables $R$ and $u=r^{\alpha}$. Here, $\alpha=1+2q/3$, where the first 
non-vanishing derivative in the Taylor expansion (6) for the density is the 
$q$th one. The central singularity is the point $R=0, u=0$. In terms of 
these variables, the geodesic equation is the Eqn. (25) of [7]:
\be
    {dR\over du} = \left( 1 - \sqrt{\Lambda\over X}\right)
    {H(X,u)\over \alpha}\equiv U(X,u).
\ee
Here, $\Lambda(r)=F(r)/r^{\alpha}$, $X=R/r^{\alpha}$. The function $H(X,u)$
is defined in [7] - we will need its form only in the limit $r\rightarrow 0$,
that limiting expression is reproduced below. The singularity turns out
to be naked if the equation $V(X_{0})=0$ 
admits one or more positive real roots $X_{0}$, where
\be
V(X) = U(X,0)-X=\left( 1 - \sqrt{\Lambda_{0}\over X}\right)
    {H(X,0)\over \alpha} - X.
\ee
We have denoted $\Lambda(0)$ as $\Lambda_{0}$. The quantity $H(X,0)$ is
given in [9] as $ H(X,0) = X + \Theta_{0}/ \sqrt{X} \equiv H_{0}$
where $\Theta_{0} = -qF_{q}/3F_{0}$, and $F_{q}$ is defined in Eqn. (8).

We are interested in finding out the nature of geodesics for the weak and
strong naked singularities. For this purpose, given a positive root to
the equation $V(X_{0})=0$, we integrate Eqn. (18) as follows. We can write
this equation as
\be
   {dX\over du} = {1\over u}\left( {dR\over du} - X\right)
   = {{U(X,u)-X}\over u}.
\ee 
The geodesics will be given as solutions $X=X(u)$ of this equation.
Given a root $X_{0}$ of $V(X_0)=0$, we can write 
$V(X) = (X-X_{0})(h_{0}-1)+h(X)$
where $h_{0}$ is a constant defined by 
$ h_{0} = (dU/dX)_{X=X_{0}}$
and the function $h(X)$ contains higher order terms in $(X-X_{0})$, that
is, $h(X_{0}) = ( dh/dX)_{X=X_{0}} = 0$.
The constant $h_{0}$ can be evaluated using (19) and is seen to be
\be
  h_{0} = {\Lambda_{0}^{1/2} H_{0}\over 2\alpha X_{0}^{3/2}}
          + {1\over \alpha} \left( 1 - \sqrt{\Lambda_{0}\over X_{0}}\right)
  \left(1 - {\Theta_{0}\over 2X_{0}^{3/2}}\right).
\ee
Eqn. (20) can be written 
\be
 {dX\over du} - (X-X_{0}) {(h_{0}-1)\over u} = {S\over u}
\ee
where we have defined $S(X,u) = U(X,u) -U(X,0) + h(X)$, and
$S(X_{0},0)=0$. This equation can be integrated to get the solution
\be
     X-X_{0}= Du^{h_{0}-1} + u^{h_{0}-1}\int S u^{-h_{0}} du,
\ee
$D$ being a constant of integration that labels different geodesics.
Note that the last term in this equation always vanishes as 
$X\rightarrow X_{0}$, $u\rightarrow 0$, irrespective of the value of $h_{0}$.
The first term, $Du^{h_{0}}-1$, vanishes in this limit if $h_{0}>1$, goes
to a constant for $h_{0}=1$ and diverges for $h_{0}<1$.

Thus, if $h_{0}>1$, a family of geodesics, labelled by the parameter $D$,
will terminate at the singularity with the root $X_{0}$ as their tangent.
The situation is different for $h_{0}\leq 1$. Consider first the range
$0<h_{0}\leq 1$. Now, as $u\rightarrow 0$, only one geodesic, labelled by
$D=0$, will terminate at the singularity with $X_{0}$ as tangent. There
will however be a family of geodesics labelled by $D\neq 0$, for which
$X\rightarrow \infty$ as $u\rightarrow 0$. These correspond to geodesics
having the $R$-axis as their tangent, rather than the root $X_{0}$
as tangent. If $h_{0}\leq 0$, then by writing the solution (23) near the 
singularity as $ R - X_{0}u = Du^{h_{0}}$
we see there will be no geodesics terminating at $R=0, u=0$.      

We now work out the classification of null geodesics families for various 
initial data configurations - this classification will clearly depend on the 
value of $h_{0}$, which can be worked out from Eqn. (21). Consider first the 
density
profile for which $\rho_{1}\neq 0$. In this case we have $q=1, \alpha=5/3$,
$\Lambda_{0}=0$. Using the result from [9] that $V(X_0)=0$ has one
positive root, given by
$X_{0}^{3/2}=\Theta_{0}/(\alpha -1)$ in (21), we get $h_{0}=2/5$.
For the case of the density profile $\rho_{1}=0$, $\rho_{2}\neq 0$,
similar considerations give $h_{0}=1/7$. Thus for both these profiles,
which result in a weak naked singularity, we have $0<h_{0}<1$. Hence
there will be one geodesic having the root $X_{0}$ as tangent, and an entire
family having the $R$-axis as the limiting tangent.

The density profile $\rho_{1}=\rho_{2}=0$, $\rho_{3}<0$ leads to a naked
singularity if the parameter $\xi$ defined earlier lies in a certain
range, as described in Section 1. Further it is a strong curvature naked
singularity. In this case, the expression for $h_{0}$ is not as trivial as 
we found above for the case of the weak naked singularity. We have
$q=3$, $\alpha=3$ and with some effort it can be shown from (21) that
\be
     h_{0} = -{1\over 6x^{4}} \left( \xi + 2x^{3}\right).
\ee
Here $X_{0}=F_{0}x^{2}$, and we have used the result from [9] that $x$
satisfies the quartic equation
\be
   2x^{4} + x^{3} + \xi x - \xi = 0.
\ee
Eqn. (24) admits values for $h_{0}$ which are greater than unity.
For instance, $x=1.1$, $\xi=-42.6$ satisfies (25) and gives $h_{0}=4.5$
from (24). Further, we know from [9] that for this density profile, whenever
the singularity is naked, there are two positive real roots. Since
$h_{0}-1$ is the value of the derivative $dV/dX$ at $X=X_{0}$, clearly
the derivative will be negative at one root and positive at the other.
Hence one of the roots, say $X_{1}$, leads to $h_{0}>1$, and the other
root, say $X_{2}$, gives $h_{0}<1$. As a result, there will be an entire
family of outgoing geodesics with $X_{1}$ as the tangent, only one geodesic
with $X_{2}$ as tangent, and a family of geodesics with the $R$-axis as
tangent. This brings out the difference in structure of geodesic families 
coming out for both the cases of weak and strong naked singularities.
For completeness we mention that the central naked singularity is a null
singularity, since it occurs at one instant of time and at one point in
space.

\bigskip

\noindent{\bf Acknowledgements}

\noindent We would like to thank I. H. Dwivedi, Kerri Newman, Amos Ori, 
Ed Siedel, Reza Tavakol and several other participants of the Pune Workshop 
on `Gravitational Collapse and Cosmic Censorship' (Dec. 1995) for useful 
discussions on some of the issues considered here. It is also a pleasure
to thank Louis Witten for helpful comments on the nature of allowed
initial data.
\bigskip

\centerline{\bf REFERENCES}
\smallskip
\begin{description}
\item[{[1]}] D. M. Eardley and L. Smarr, Phys. Rev. D {\bf 19} (1979) 2239.
\item[{[2]}] D. Christodoulou, Commun. Math. Phys. {\bf 93} (1984) 171.
\item[{[3]}] R. P. A. C. Newman, Class. Quantum Grav. {\bf 3} (1986) 527.
\item[{[4]}] B. Waugh and K. Lake, Phys. Rev. D {\bf 38} (1988) 1315.
\item[{[5]}] G. Grillo, Class. Quantum Grav. {\bf 8} (1991) 739. 
\item[{[6]}] I. H. Dwivedi and P. S. Joshi, Class. Quantum Grav.
{\bf 9} (1992) L69.
\item[{[7]}] P. S. Joshi and I. H. Dwivedi, Phys. Rev. D {\bf 47} (1993) 5357.
\item[{[8]}] P. S. Joshi and T. P. Singh, Phys. Rev. D51 (1995) 6778.
\item[{[9]}] T. P. Singh and P. S. Joshi, Class. Quantum Grav. 
{\bf13} (1996) 559.
\item[{[10]}] R. C. Tolman, Proc. Natl. Acad. Sci. USA {\bf 20} (1934) 410.
\item[{[11]}] H. Bondi, Mon. Not.  Astron. Soc. {\bf107} (1947) 343.
\item[{[12]}] C. S. Unnikrishnan, Phys. Rev. {\bf D53} (1996) R580.
\item[{[13]}] S. Chandrasekhar, An Introduction to the Study of Stellar
Structure (1958), Dover Edition, Chapter IV, Section 5.
\item[{[14]}] L. Spitzer, Dynamical Evolution of Globular Clusters (1987)
Princeton University Press; Section 1.1
\item[{[15]}] B. J. Carr and S. W. Hawking, Mon. Not. R. Astr. Soc.
{\bf 168} (1974) 399; R. N. Henriksen and Paul S. Wesson,
Astrophysics and Space Science {\bf 53} (1978) 429.
\item[{[16]}] Amos Ori and Tsvi Piran, Phys. Rev. {\bf D42} (1990) 1068.
\item[{[17]}] J. Oppenheimer and H. Snyder, Phys. Rev. {\bf 56} (1939) 455.
\item[{[18]}] H. M. Antia, Phys. Rev. {\bf D53} (1996) 3472.
\item[{[19]}] I. H Dwivedi and P. S. Joshi, 
Class. Quantum Grav. {\bf 6} (1989) 1599; 
Class. Quantum Grav. {\bf 8} (1991) 1339; 
P. S. Joshi and I. H. Dwivedi, Gen. Rel. and Gravn. {\bf24} (1992) 129.
\item[{[20]}] See e.g. Section V of Ref 7;  
S. A. Hayward, Phys. Rev. {\bf D53} (1996) 1938. 
\item[{[21]}] S. L. Shapiro and S. A. Teukolsky, Phys. Rev. Lett.
{\bf 66} (1991) 994.

\end{description}

\end{document}